\documentclass[aps,prx,twocolumn,nofootinbib]{revtex4-2}
\usepackage{graphicx}% Include figure files
\usepackage{dcolumn}% Align table columns on decimal point
\usepackage{bm}% bold math
\usepackage{ textcomp }
\usepackage{color}
\usepackage{amsmath}
\usepackage{amssymb}
\usepackage{xcolor}
\usepackage{hyperref}
\usepackage{easyReview}
\usepackage{mathdots}
\usepackage{float}
\usepackage{lineno}
\usepackage{array}
\usepackage{xcolor,colortbl}
\usepackage[normalem]{ulem}
\definecolor{LightBlue}{rgb}{0.8,0.8,0.8}
\allowdisplaybreaks
\usepackage{amsmath}

\newtheorem{Proposition}{Proposition}

\begin{document}
\title{Quantum-inspired activation functions and quantum Chebyshev-polynomial network}
\author{Shaozhi Li}
\email[Corresponding author: ]{shaozhl@clemson.edu}
\affiliation{National Center for Transportation Cybersecurity and Resiliency (TracR), Clemson University, SC 29631, USA}

\author{M Sabbir Salek}
\affiliation{National Center for Transportation Cybersecurity and Resiliency (TracR), Clemson University, SC 29631, USA}

\author{Yao Wang}
\affiliation{Department  of Chemistry, Emory University, Atlanta, GA 30322, USA}

\author{Mashrur Chowdhury}
\affiliation{National Center for Transportation Cybersecurity and Resiliency (TracR), Clemson University, SC 29631, USA}
\affiliation{Glenn Department of Civil Engineering, Clemson University, Clemson, SC 29634, USA}

%\renewcommand{\thefootnote}{\fnsymbol{footnote}} % Use symbols for footnote markers
%\footnotetext[1]{Corresponding author. Email: shaozhl@clemson.edu}

\begin{abstract}
Driven by the significant advantages offered by quantum computing, research in quantum machine learning has increased in recent years. While quantum speed-up has been demonstrated in some applications of quantum machine learning, a comprehensive understanding of its underlying mechanisms for improved performance remains elusive. Our study address this problem by investigating the functional expressibility of quantum circuits integrated within a convolutional neural network (CNN). Through numerical experiments on the MNIST, Fashion MNIST, and Letter datasets, our hybrid quantum-classical CNN model demonstrates superior feature selection capabilities and substantially reduces the required training steps compared to classical CNNs. Notably, we observe similar performance improvements when incorporating three other quantum-inspired activation functions in classical neural networks, indicating the benefits of adopting quantum-inspired activation functions. Additionally, we developed a hybrid quantum Chebyshev-polynomial network (QCPN) based on the properties of quantum activation functions. We demonstrate that a three-layer QCPN can approximate any continuous function, a feat not achievable by a standard three-layer classical neural network. Our findings suggest that quantum-inspired activation functions can reduce model depth while maintaining high learning capability, making them a promising approach for optimizing large-scale machine-learning models. We also outline future research directions for leveraging quantum advantages in machine learning, aiming to unlock further potential in this rapidly evolving field.
\end{abstract}

\maketitle

\section{Introduction}
The unparalleled potential of quantum algorithms over their classical counterparts has ignited widespread enthusiasm for quantum computing~\cite{LaddNature2010,CaoCR2019,LiuNP2021,aaronson2016complexitytheoretic,KlcoPRA2020,NiuPRL2022,JeremyScience,Jerbi2023NC,SoodIEEE2023,PainePRA2023}. 
Due to achievements in quantum hardware development~\cite{KokRMP2007,BarendsNature2014,BallancePRL2016,IuliaNRP2020,EveredNature2023}, quantum supremacy was demonstrated via the random sampling task made by Google superconducting quantum computers and photonic quantum devices~\cite{AruteNature2019,ZhongScience2020}.
In addition to these landmark achievements, quantum computing has found applications across diverse domains, including simulating many-body Hamiltonian~\cite{PeruzzoNC2014,MalleyPRX2016,KivlichanPRL2018,tubman2018postponing,CaoCR2019,DongPRX2022,qing2023use,AlistairQS2023}, simulating spectroscopies~\cite{BauerPRX2016,BakerPRA2021,JacopoPRR2022,QiaoyiPRR2022,SantosPRL2023}, and solving NP-complete problems~\cite{farhi2001quantum,Arrazolanpj2018,Centrone2021,ZhangLSA2021}. Among them, most of these applications use the variational technique to find a solution with a minimal value of the loss function or the energy, which is analogous to the strategy used in machine learning techniques, inspiring the idea of quantum machine learning~\cite{BiamonteNature2017,ciliberto2018}.

The current extensively used supervised quantum machine learning methods use quantum models or circuits to extract essential features~\cite{schuld2021supervised,Blanknpj2020,Zoufainpj2019,Huangnpj,TancaraPRA2023,SlatteryPRA2023}. Unlike the kernel method used in classical machine learning, the quantum model encodes data in the high-dimensional Hilbert space and utilizes a series of quantum gates to construct nonlinear quantum kernel functions. 
Broadly, these quantum kernel methods fall into two categories: explicit and implicit quantum models~\cite{HavlicekNature2019}. 
Explicit models utilize measurements of a quantum state controlled by encoded data and parameterized gates.
The recently proposed data reuploading model also belongs to this category~\cite{AdrianQuantum2020,Moreira2023npj}. Implicit models weigh the inner products of quantum states controlled by the encoded input data. A representative example of the implicit model is the quantum support vector machine~\cite{RebentrostPRL2014,lloyd2020quantum,peter2021npj,Kusumoto2021,JagerNC2023}. Compared to implicit models, explicit models create more complex functions, posing significant challenges to our understanding, which in turn obscures the mechanisms underpinning their superior performance in some case studies
~\cite{Alam2021IEEE,Trochun2021IEEE,NakajiPRR2022,AjlouniBMC2023,Hasan2023,Zeguendry2023,Rath2023QuantumDE}.

Understanding the origin of superior performance in quantum machine learning is essential for optimizing quantum models and potentially enriching classical algorithms. Quantum-inspired classical algorithms, for example, have already accelerated linear algebra computations, showcasing the benefits of translating quantum insights to classical contexts~\cite{gilyen2018quantuminspired,chia2018quantuminspired,Arrazola2020,EwinPRL2021}. 
Despite these advancements, the machine learning field sees a paucity of quantum-inspired algorithms, primarily due to an inadequate grasp of quantum model intricacies~\cite{Hwang2024Quantum}.
This knowledge gap has spurred investigations into the expressibility of quantum models~\cite{SimAQT2019,SchuldPRA2020, DuPRR2020,Wu2024Randomness}.
However, uncovering the relationship between the expressibility and the architecture of a quantum circuit is challenging. Consequently, research has predominantly pursued numerical evaluations, including evaluating the expressibility of the Hilbert space and the functional expressibility~\cite{SimAQT2019,SchuldPRA2020, DuPRR2020,Wu2024Randomness}. For instance, the expressibility of the Hilbert space was characterized by comparing different distributions of the fidelity between the quantum state from a quantum model and the Haar-random states~\cite{SimAQT2019}. In addition, Schude et al. proposed utilizing Fourier series to explain the functional expressibility of a quantum model~\cite{SchuldPRA2020}. While these insights advance our understanding, they have not elucidated the relationship between Hilbert space expressibility and quantum machine learning efficacy. 
Moreover, the decomposition of a function into a Fourier series cannot intuitively reflect the nonlinear relationship between the input data and the output. To advance quantum machine learning techniques, it is vital to address these foundational questions.

Our research advances the understanding of quantum models in quantum machine learning by providing clear analytical expressions for quantum circuits, distinguishing our work from previous studies. Specifically, we investigate two shallow quantum circuits integrated within a convolutional neural network (CNN) framework. Compared to classical CNNs, the hybrid quantum-classical CNN model requires significantly fewer training steps while achieving comparable training accuracy. Notably, we demonstrate that a special quantum activation function can efficiently extract edge information from an object while filtering out redundant data, resulting in faster convergence of the loss function. Additionally, we explore three quantum-inspired activation functions, all of which achieve high accuracy with fewer training steps compared to classical CNNs. These findings suggest that the higher-order nonlinear properties of quantum-inspired activation functions can enhance the performance of machine learning models. Furthermore, we develop a hybrid quantum Chebyshev-polynomial network (QCPN), capable of approximating any continuous function using only three neural network layers. In contrast, a standard three-layer classical neural network cannot achieve this level of approximation. This result highlights the potential of quantum-inspired activation functions to reduce the depth of machine learning models, offering a promising approach for optimizing large-scale models.

This paper is organized as follows. We highlight the major contributions and the importance of our work in Sec.~\ref{sec:I}.
In Sec.\ref{sec:II}, we present the functional formulation of a generic quantum circuit. In Sec.\ref{sec:III}, we analyze the training results of a hybrid quantum-classical CNN using three distinct datasets. We also introduce three quantum-inspired activation functions and evaluate the performance of the CNN with these functions. Furthermore, we develop a quantum Chebyshev-polynomial network, capable of approximating any continuous function using three layers. Finally, in Sec.~\ref{sec:IV}, we discuss the future directions of quantum machine learning and provide perspectives on its ongoing development.

\begin{figure}[h]
\centering
\includegraphics[width=0.99\columnwidth]{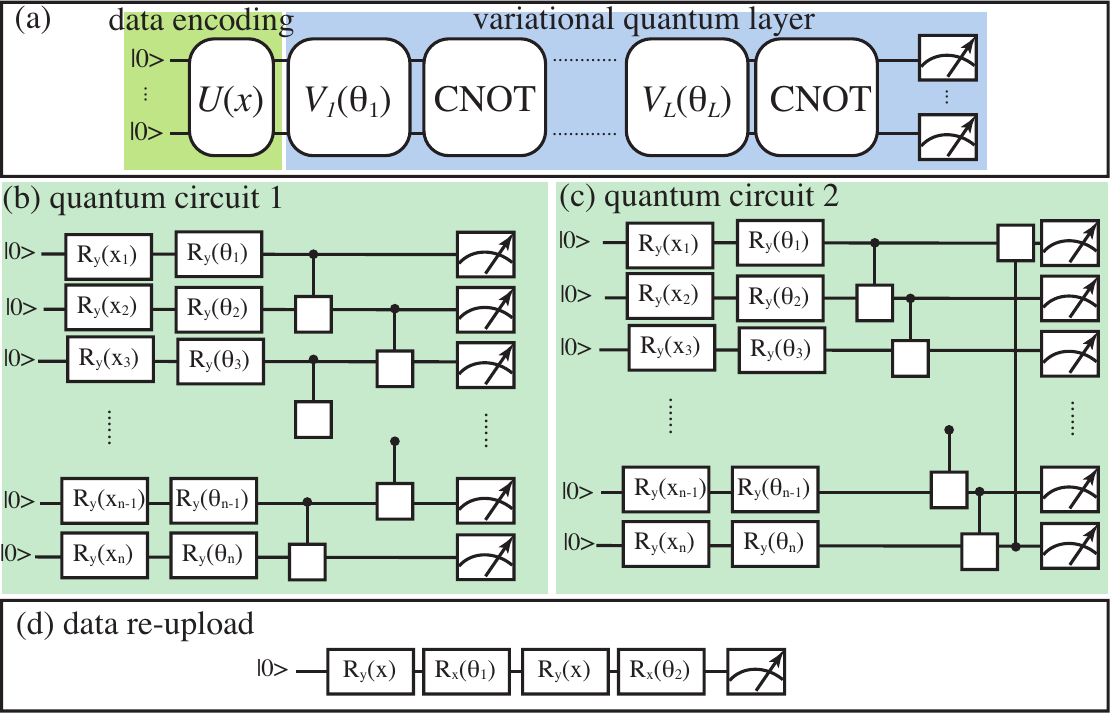}
\caption{(a) Schematic of a generic quantum circuit, consisting of a data-encoding layer and a variational quantum layer. (b) Architecture of Quantum Circuit 1. (c) Architecture of Quantum Circuit 2. (d) One-qubit circuit used for data re-uploading.
}
\label{Fig:fig1}
\end{figure}

\section{Contribution}\label{sec:I}
The major contributions of our work include the following three aspects:
\begin{itemize}
\item We explained the functional expression of a generic quantum circuit that can be integrated into the classical machine learning methods. This explanation is essential to our understanding of quantum machine learning techniques. Furthermore, our work opens a new research direction on exploring quantum-inspired classical activation functions.
\item We discovered an interesting quantum activation function, which can extract edge information and drop redundant information inside an object.
\item We developed a quantum Chebyshev-polynomial network, which can approximate any continuous function using three neural network layers. However, a classical three-layer neural network does not have this learnability. Therefore, our developed quantum Chebyshev-polynomial network could be adopted to reduce the complexity of deep learning models while maintaining the same learning capability.
\end{itemize}

\section{Functional Expressibility}\label{sec:II}
To understand the mechanism of quantum neural networks, we analyze the functional expressibility of a generic quantum circuit defined in Fig.~\ref{Fig:fig1} (a). As a linear quantum model, we can divide it into two parts: data encoding and variational quantum layers. In the data encoding part, we transform real-world data ${\bf x}$ to a quantum state using quantum operator $U({\bf x})$, which consists of rotation gates. The variational quantum layer consists of $L$ rotation layers ($V_l({\bf \theta})$) and $L$ CNOT layers. By applying measurements with Pauli operators $M$, the output $O$ of this quantum circuit is
\begin{equation}\label{eq: 1}
O=\langle 0 | U^{\dagger}({\bf x})V^{\dagger}({\bf \theta}_1,{\bf \theta}_2,\cdots,{\bf \theta}_L) M V({\bf \theta}_1,{\bf \theta}_2,\cdots,{\bf \theta}_L) U({\bf x}) |0\rangle
\end{equation}
where $V({\bf \theta}_1,{\bf \theta}_2,\cdots,{\bf \theta}_L)$ denotes the quantum operator in the variational circuit.

By further analyzing the matrix element of $U(\bf x)$, $V({\bf \theta}_1,{\bf \theta}_2,\cdots,{\bf \theta}_L)$, and $M$, we find that $O$ can be rewritten as
\begin{align}
O=&\sum_{s}f_s({\bf \theta},{\bf x}),\label{eq: 2}\\
f_s({\bf \theta},{\bf x})=&c_s\prod_{i\in \mathcal{L}_{s,1}} \mathrm{cos}(\theta_i/2+q_i) \prod_{j \in \mathcal{L}_{s,2}} \mathrm{cos}(\theta_j/2+q_j) \times \nonumber\\
&\prod_{k \in \mathcal{L}_{s,3}} \mathrm{cos}(\pi x_k/2+q_k) \prod_{l \in \mathcal{L}_{s,4}} \mathrm{cos}(\pi x_l/2+q_l).\nonumber\\
\end{align}
Here, $\mathcal{L}_\gamma$ is a set of integers, and the value of $c_s$ is either 1 or -1. $q_i$ belongs to the group of \{$0, \pi/2$\}. Equation~(\ref{eq: 2}) implies that the output of a quantum circuit is a polynomial function whose component is a continued product of the cosine series. 

To clarify this polynomial function, we investigate two quantum circuits as examples, which are plotted in Fig.~\ref{Fig:fig1} (b) and  Fig.~\ref{Fig:fig1} (c). In the quantum circuit 1 (QC 1) shown in Fig.~\ref{Fig:fig1} (b), the data encoding is achieved via $R_y$ gates, and the variational quantum circuit is made of one $R_y$ rotation layer and one CNOT layer, in which the CNOT gate is applied between two adjacent qubits, which first across even-indexed qubits and then odd-indexed qubits. In the quantum circuit 2 (QC 2) shown in  Fig.~\ref{Fig:fig1} (c), the CNOT layer is replaced by a layer that applies a CNOT gate between each two adjacent qubits. With some derivations (see details in the Appendix), it is found that the output of QC 1 can be written as
\begin{align}\label{eq:QC1}
O_{QC1}(\theta,x)=\left\{ \begin{matrix}
-\prod^n_{i\in\text{odd}} \text{cos}\alpha_i & n\in\text{odd} \\
-\text{cos}\alpha_n\prod_{i\in\text{odd}}^{n-3} \text{cos}\alpha_i & n\in\text{even}
\end{matrix}
\right.
\end{align}
where $\alpha_i=\theta_i + \pi x_i$, and $n$ is the number of qubits. Additionally, the output of QC 2 can be expressed as
\begin{widetext}
\begin{equation}\label{eq: 3}
O_{QC2}(\theta,x) = \left\{ \begin{matrix}
\left(\prod_{i=n-2}^{i=i-4,i>4}\text{cos}\alpha_{i}\right)\text{cos}\alpha_3\text{cos}\alpha_1 & \text{when } n=4+4m\nonumber\\
\left(\prod_{i=n-2}^{i=i-4,i>6}\text{cos}\alpha_{i}\right)\text{cos}\alpha_4\text{cos}\alpha_2 & \text{when } n=6+4m\nonumber\\
\left(\prod_{i=n-2}^{i=i-4,i>5}\text{cos}\alpha_{i}\right)\text{cos}\alpha_5\text{cos}\alpha_2 & \text{when }  n=3+4m\nonumber\\
\left(\prod_{i=n-2}^{i=i-4,i>5}\text{cos}\alpha_{i}\right)\text{cos}\alpha_3\text{cos}\alpha_1 & \text{when } n=5+4m\nonumber
\end{matrix} \right.
\end{equation}
\end{widetext}
where $m$ is a non-negative integer.

As observed from the equations above, the polynomial function transitions into a monotonic function for QC 1 and QC 2. In general, increasing the number of rotation layers in the variational quantum layers enhances the complexity of the output function by adding more components. This conclusion also applies to the data re-uploading technique illustrated in Fig.~\ref{Fig:fig1}(d). In this method, the repeated application of rotation gates effectively increases the number of matrix multiplications, which, in turn, augments the number of components in the resulting polynomial function.

\begin{figure*}[ht]
\centering
\includegraphics[width=0.99\textwidth]{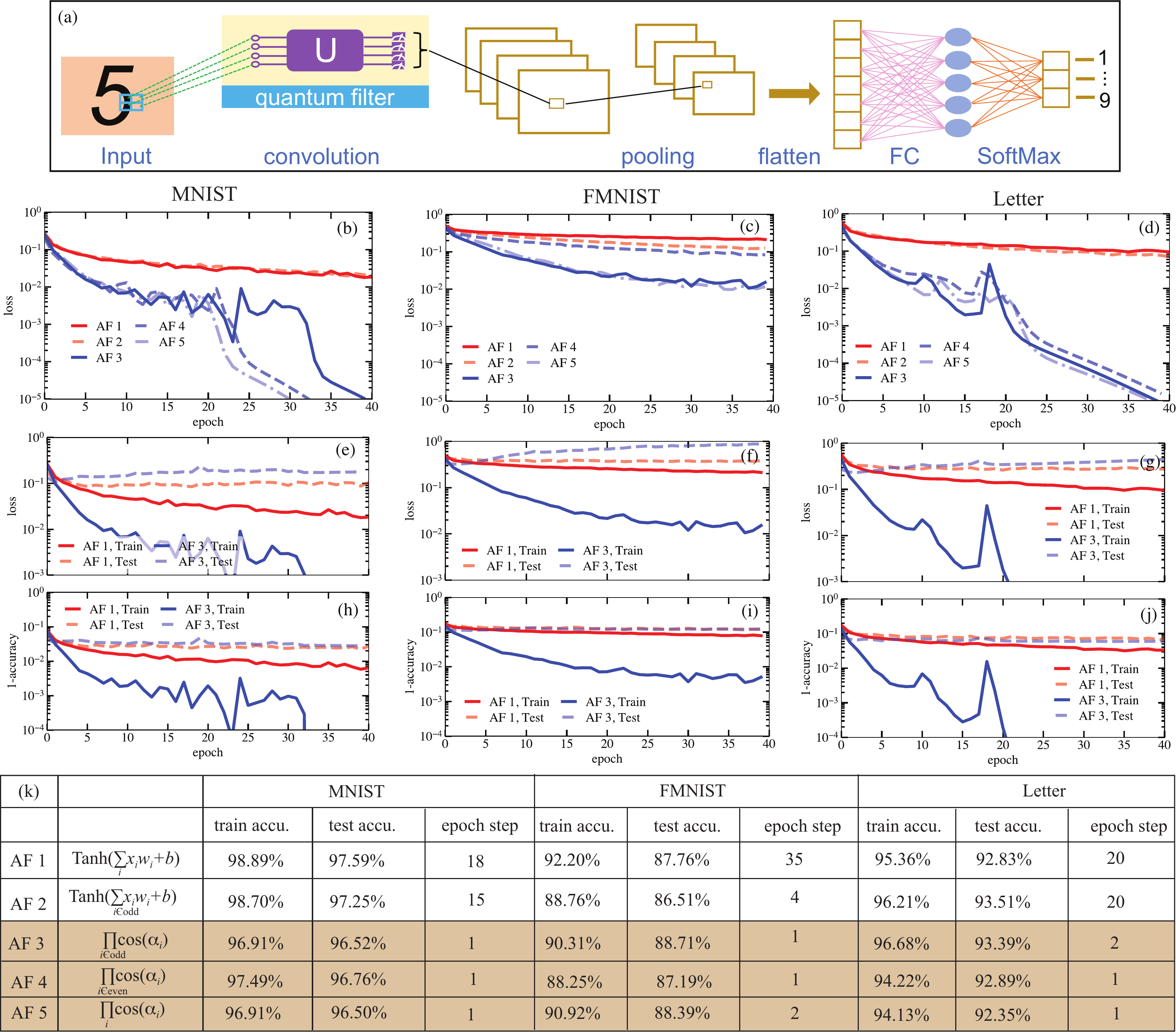}
\caption{(a) Hybrid quantum-classical convolutional neural network (CNN) architecture. (b)-(d) Training loss curves for the CNN using various activation functions across epochs on the MNIST, FMNIST, and Letter datasets, respectively. (e)-(g) Comparison of the training and testing loss curves for the CNN using Activation Functions (AF) 1 and 3 across epochs for the MNIST, FMNIST, and Letter datasets, respectively. (h)-(j) Accuracy residuals for the CNN using AF 1 and AF 3 across epochs for both training and testing on the MNIST, FMNIST, and Letter datasets. (k) Summary table of the training accuracy, testing accuracy, and training steps for CNNs using different activation functions.
}
\label{Fig:fig2}
\end{figure*}

\begin{figure*}[ht]
\centering
\includegraphics[width=0.99\textwidth]{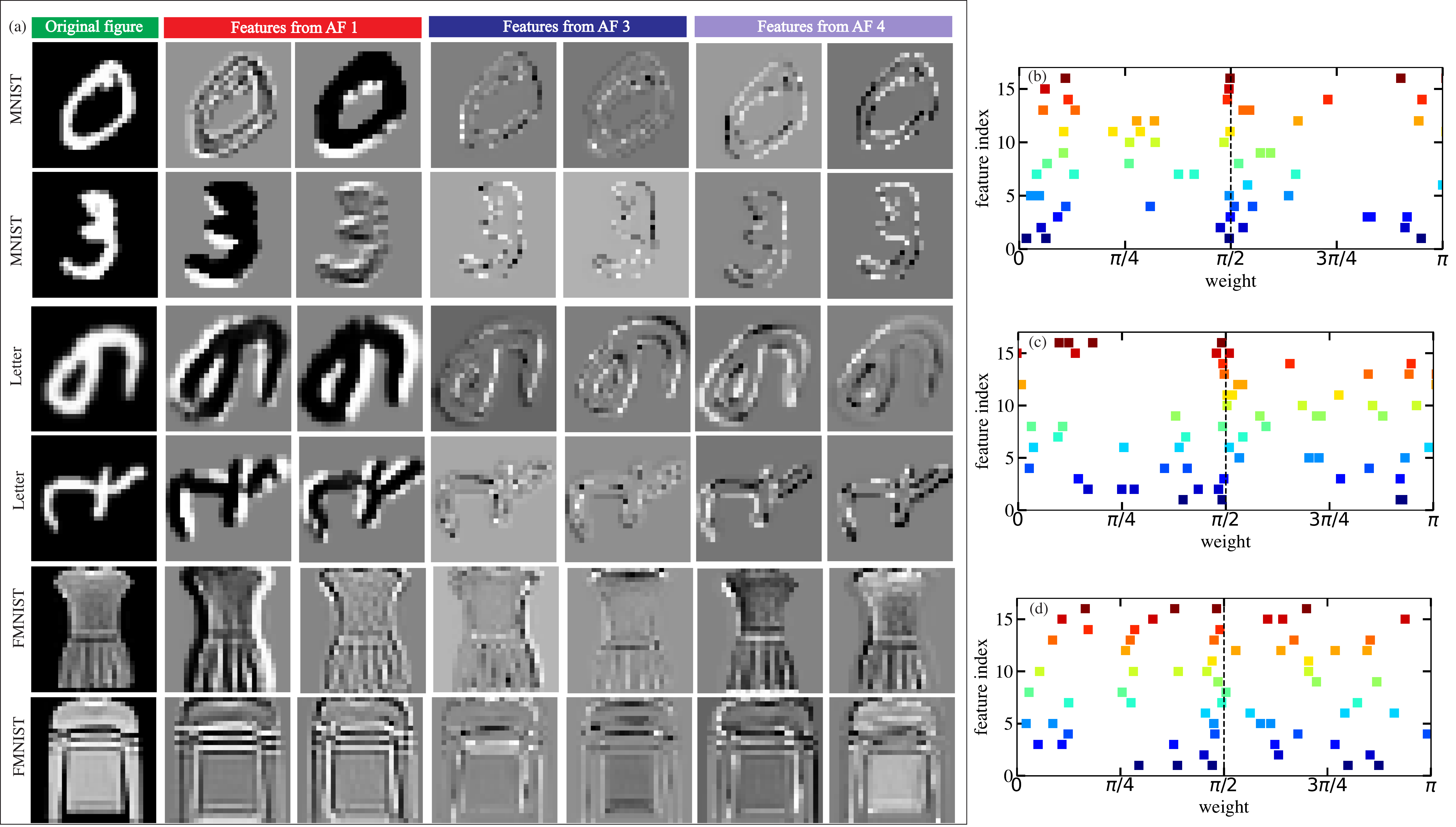}
\caption{(a) The original input image and the corresponding feature maps in the first hidden layer of the convolutional neural network (CNN) using Activation Functions (AF) 1, 3, and 4, as shown in Fig. 2(k). (b)-(d) Trained angle parameters of the CNN with AF 3 for the MNIST, FMNIST, and Letter datasets, respectively. Different colors along the vertical direction denotes different features.
}
\label{Fig:fig3}
\end{figure*}

\section{Quantum neural network}\label{sec:III}
Quantum neural networks can be employed to perform various machine-learning tasks. However, their application to real-world problems is constrained by the limited availability of quantum resources. As a result, using purely quantum neural networks to solve problems involving high-dimensional data is currently impractical. To overcome this challenge, two primary strategies are often utilized. The first approach involves hybrid quantum-classical CNNs, which only require a small patch of an image to be processed by the quantum model. The second approach reduces the dimensionality of the input data using classical machine learning techniques such as principal component analysis (PCA) or autoencoders. In this situation, a pure quantum model can be adopted to solve the problem. However, we note that a simple classical two-layer neural network can be easily trained with high accuracy using these extracted features. It seems that adopting quantum models is not necessary for the second strategy. Therefore, our work will focus on the first strategy.

\subsection{Hybrid quantum-classical convolutional neural network}
In the hybrid quantum-classical CNN, the quantum model is adopted in the convolutional operation, as shown in Fig.~\ref{Fig:fig2} (a), which includes one convolutional layer with quantum filters, one average pooling layer, one flattened layer, and two fully connected (FC) neural networks (NN). For the output layer, a SoftMax activation function is employed to facilitate multi-class classification alongside the use of a categorical cross-entropy loss function.

In the classical CNN, the convolutional calculation is performed with Tanh or Relu activation functions. In the hybrid CNN, a polynomial function generated from the quantum model replaces this classical activation function. Considering the simplicity of the classical activation function, we assume that a simple function generated from a quantum model can make the hybrid CNN work. The simplest function of polynomial functions is the monotonic function; therefore, we adopt QC 1 in our hybrid CNN. For a nine-qubit circuit, there is no significant difference between $O_{QC1}$ and $O_{QC2}$ (see details in Appendix).
Compared to the Tanh function used in the classical CNN, $O_{QC1}$ and $O_{QC2}$ have higher-order nonlinearity and site dependence. To understand the effect of these differences on the performance of the CNN, we perform simulations using five activation functions, as shown in Fig.~\ref{Fig:fig2}(k). These activation functions include the standard Tanh function (labeled as AF 1), a site-dependent Tanh function (labeled as AF 2), $O_{QC1}$ (labeled as AF 3), a continued product of the cosine series with even indexes (labeled as AF 4), and a continued product of the cosine series without the site dependence (labeled as AF 5).

Our CNNs are adopted to classify images from three datasets: MNIST, Fashion MMNIST (FMNIST), and Letter datasets. For each dataset, 52000 samples with ten classes are used to train the CNNs, and 8000 samples are used for testing. For the CNNs, the number of channels within the first hidden layer is 16, and the number of neurons in the last-second hidden layer is 64. The kernel size is set as $3\times 3$. Figures~\ref{Fig:fig2}(b)-~\ref{Fig:fig2}(d) plot the loss function of these five CNNs for three datasets. 
Compared to the loss function of CNNs using AF 1 and AF 2, the loss functions of CNNs with AF 3 and AF 5 can efficiently converge to a smaller value, implying that the quantum activation functions have higher learnability than the classical activation functions. However, this advantage is not observed in the CNN using AF 4 for the FMNIST dataset, implying that AF 2 is not an optimal activation function for the FMNIST dataset.

We further evaluate the performance of CNNs using AF 1 and AF 3 by analyzing the loss functions and accuracy metrics for both the training and testing datasets, as shown in Figs.~\ref{Fig:fig2}(e)–~\ref{Fig:fig2}(j). Although the quantum activation functions exhibit higher learnability, the testing accuracy of CNNs with AF 3 is comparable to that of the standard classical CNNs. We define the optimal model by selecting the epoch corresponding to the minimum loss value on the testing dataset. The table at the bottom of Fig.~\ref{Fig:fig2} summarizes the training and testing accuracies, along with the optimal epoch for each of these five CNNs. Compared to the standard classical CNNs, the optimal hybrid CNN models can be obtained with much fewer epochs while achieving similar testing accuracies. These results imply that adopting the quantum activation function can make machine learning more efficient.

It was said that quantum machine learning selects features in the high-dimensional Hilbert space, which has made it shrouded in mystery. However, we note that feature constructions are made by measurements; therefore, understanding feature selections based on measurements is more straightforward than analyzing quantum states. Here, we clarify feature selections using quantum activation functions. Figure~\ref{Fig:fig3}(a) displays features of trained CNNs using AF 1, AF 3, and AF 4 in the first hidden layer for MNIST, FMNIST, and Letter datasets. Compared to the results of the standard classical CNN, CNNs using quantum activation functions exhibit significantly different features. For the MNIST and the Letter datasets, the model with AF 1 captures comprehensive detail, including both edge delineation and internal patterns within the digit imagery. In contrast, the model leveraging the cosine functions ({\it i.e.} AF 3 and AF 4) primarily emphasizes edge information. For the FMNIST dataset, the information inside an object of CNNs using AF 3 and AF 4 does not disappear but is less than that of the CNN using AF 1.

To elucidate this different feature selection, we explore the function  $f({\bf \alpha})=\prod_i^3 \text{cos}\alpha_i$, where $\alpha_i=\theta_i+\phi_i$. In this formula, $\theta_i$ denotes the angle corresponding to the encoded data, and $\phi_i$ represents a variable. By considering a three-pixel patch, encoding the background pixels of an image as zero yields an output of zero whenever any $\phi_i$ equals $\pi/2$ or $-\pi/2$. This zero output is similarly achieved when encoding a patch inside a digital imagery as $\pi$.
At a smooth edge, where $\theta_i$ lies between 0 and $\pi$, $f({\bf \alpha})$ is nonzero for $\phi=\pm\pi/2$.
Therefore, $f({\bf \alpha})$ can distinguish the edge information and drop redundant information by setting $\phi_i=\pm\pi/2$. We examined the trained weight $\phi_i$ of the CNN with AF 3 for MNIST, FMNIST, and Letter datasets, plotted in Figs.~\ref{Fig:fig3}(b)-~\ref{Fig:fig3}(d). Different colors denote different features. It is found that the weight of each feature has an element close to $\pi/2$. This result demonstrates that our specific quantum activation has the ability to identify the edge information of an object. This ability could disappear if a more complicated quantum activation function is used. 
With this understanding, we can now clarify the existence of nonzero information inside an object for the FMNIST dataset. The reason is that there is no pixel with a value of zero inside an object; therefore, the value of $\mathrm{cos}(\theta_i+\pi/2)$ is no longer zero.

\subsection{Quantum-inspired activation functions}
\begin{figure}[t]
\centering
\includegraphics[width=0.99\columnwidth]{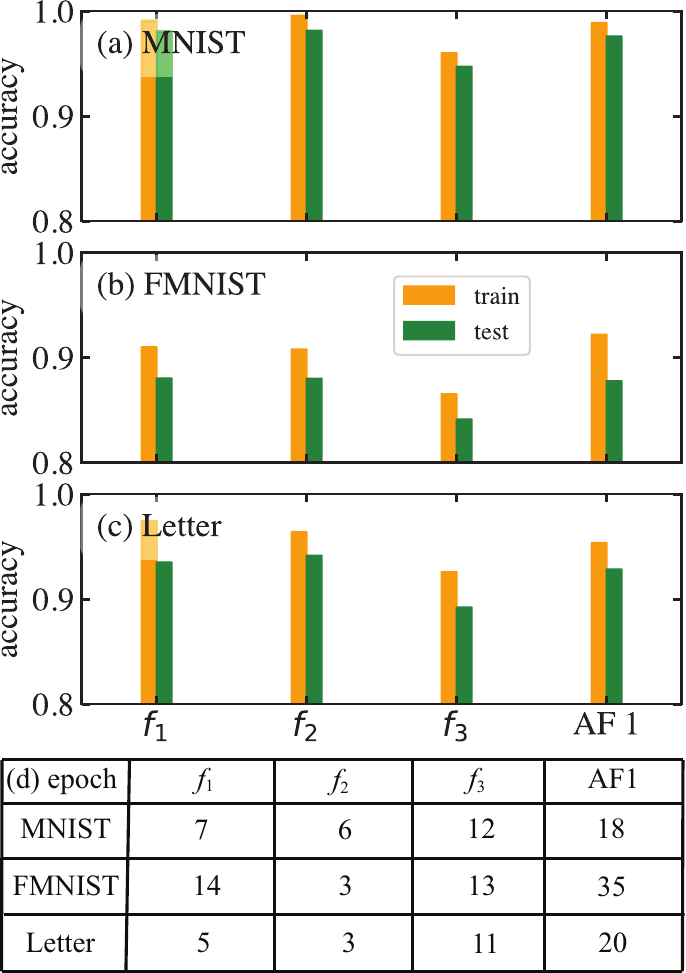}
\caption{(a)-(c) Accuracy of the convolutional neural network (CNN) using activation functions $f_1$, $f_2$, $f_3$, and AF 1 for the MNIST, FMNIST, and Letter datasets, respectively. (d) Table summarizing the training steps of these four CNNs across these three datasets.
}
\label{Fig:fig4}
\end{figure}

\begin{figure*}[t]
\centering
\includegraphics[width=0.99\textwidth]{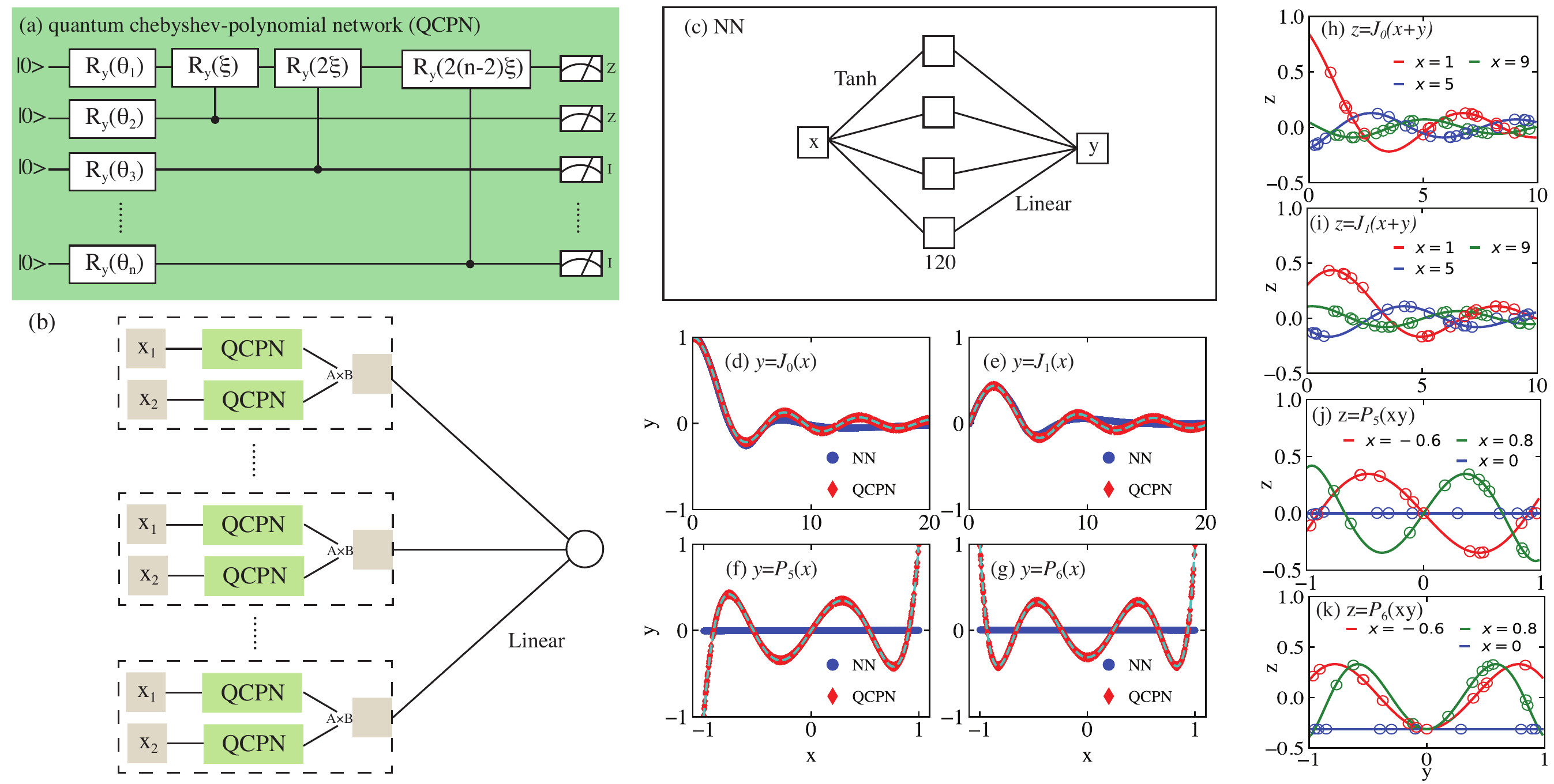}
\caption{(a) Architecture of the quantum Chebyshev-polynomial network (QCPN). (b) Hybrid QCPN architecture for a two-element input. (c) Three-layer neural network (NN) architecture. (d)-(g) Predictions of the hybrid QCPN model and NN model for four special functions: $y=J_0(x)$, $y=J_1(x)$, $y=P_5(x)$, and $y=P_6(x)$, respectively. The dashed curves represent the analytical solutions. (h)-(k) Predictions of the hybrid QCPN model for four special functions: $z=J_0(x+y)$, $z=J_1(x+y)$, $z=P_5(xy)$, and $z=P_6(xy)$, respectively. The solid curve denotes the analytical result.
}
\label{Fig:fig5}
\end{figure*}

As we explained in Eq.~(\ref{eq: 1}), the output of a generic quantum circuit is a polynomial function in which each term is a continued product of the cosine function. In practice, it would be difficult to derive a detailed formula for a complicated quantum circuit. However, it is not necessary to derive such a formula once we understand the mechanism of the hybrid CNNs. Implementing different quantum circuits is equivalent to adopting different polynomial functions. With this understanding, we can directly implement polynomial functions in our classical computer rather than adopting a quantum circuit.

We propose three distinct polynomial functions to initiate a new research direction focused on developing efficient polynomial functions for neural networks. These functions are described as follows:
\begin{align}
f_1=&\prod_{i\in \text{even}} (\text{cos}(\phi_i)+\text{cos}(\theta_i)) + \prod_{i\in \text{odd}} (\text{sin}(\phi_i)+\text{cos}(\theta_i))\nonumber\\
&\\
f_2=&\prod_{i\in \text{even}} (\text{cos}(\phi_i)+\text{cos}(\theta_i))\\
f_3=&\prod_{i} \text{cos}(\phi_i)\text{cos}(\theta_i),
\end{align}
where $\theta$ denotes the angle corresponding to the encoded data, and $\phi$ denotes the parametric angle. We note that these three polynomial functions are not derived from any quantum circuit. 

Figures~\ref{Fig:fig4}(a)-~\ref{Fig:fig4}(c) show the accuracy across three datasets: MNIST, FMNIST, and Letter. The orange bars represent the training data, while the green bars correspond to the testing data. The CNNs utilizing the $f_1$ and $f_2$ functions achieve performance comparable to the CNN with AF 1. In contrast, the accuracy of the CNN using the $f_3$ function is slightly lower across all three datasets. Figure~\ref{Fig:fig4} (d) summarizes the number of training steps required to achieve the optimal model for each dataset. Notably, CNNs with quantum-inspired activation functions require fewer training steps compared to the classical CNN using AF 1. These findings demonstrate that manually designing polynomial functions can also support the successful training of CNNs and improve their performance.

\subsection{Quantum Chebyshev-polynomial network}
Higher-order nonlinear activation functions can not only reduce the number of training steps but also decrease the depth of deep learning models. In this section, we develop a quantum Chebyshev-polynomial network leveraging the properties of quantum activation functions to demonstrate the efficiency of higher-order nonlinear activation functions.

As is well known that any continuous function $F(x)$ defined over the interval $[a, b]$ can be approximated using a Chebyshev polynomial as follows:
\begin{equation}
F(x) = \sum_{n=0}^{\infty} c_n T_n(L(x)) +\epsilon,
\end{equation}
where $L(x)$ is a linear transformation that maps $x$ to the interval $[-1, 1]$, $T_n(\xi)$ is the Chebyshev polynomial of the first kind, and $\epsilon$ is a small error term. Alternatively, $F(x)$ can can be expanded using a Fourier series. However, the Chebyshev polynomial expansion has the advantage that the coefficients $c_n$ rapidly decrease as $n$ increases, allowing us to approximate $F(x)$ accurately with only a finite number of terms.

For a machine learning task where the target value is a continuous function of the input, we can use Chebyshev polynomials to establish a relationship between the target value and the input. Based on this concept, we develop a quantum Chebyshev-polynomial network (QCPN), as illustrated in Fig.~\ref{Fig:fig5}(a). The QCPN is constructed using parametric $R_y$ rotation gates and controlled-$R_y(\xi)$ gates, where $\xi=\mathrm{arccos}(x)$. Here, we assume the input $x$ lies within the range of [-1, 1]. The measurement is performed by applying the Pauli $Z$ gate on the first and second qubits, while the identity operator is applied to the remaining qubits. For an $n$-qubit ($n>2$) circuit, the output of the QCPN can be expressed as
\begin{equation}
O(x)=\sum_i^{(n-1)(n-2)} (-1)^i a_i^2 T_{i}(x),
\end{equation}
where $a_i$ is determined by the angle in the parametric $R_y$ gates. Since the $a_i$ values are not independent of each other, multiple QCPNs can be employed to increase the degrees of freedom. Consequently, the target value $y(x)$ can be expressed as:
\begin{equation}
y(x) = \sum_{i} w_i O_i(x) + b, \label{eq: multicheby}
\end{equation}
where $O_i$ denotes the output of the $i$-th QCPN, and $w_i$ represents the corresponding weight.

When the input consists of $m$ elements, the target value can be approximated as
\begin{equation}
y(x_1, x_2, \cdots, x_m)=\sum_i w_i \prod_j^m O_{i,j}(x_j) +b.\label{eq: multchev}
\end{equation}
If $y(x_1, x_2,\cdots, x_m)$ is a separable function, {\it i.e.} $y(x_1, x_2,\cdots, x_m)=\prod_j^m y_j(x_j)$, then Eq.~(\ref{eq: multchev}) naturally holds.
Figure~\ref{Fig:fig5}(b) illustrates the architecture of the hybrid quantum network for an input with two elements. The notation $A\times B$ in Fig.~\ref{Fig:fig5}(b) represents the multiplication of the outputs from quantum circuits with inputs $x_1$ and $x_2$. Equation~(\ref{eq: multchev}) implies that we can approximate any continuous function using a hybrid QCPN with three layers. In contrast, a standard three-layer neural network cannot achieve this level of expressivity, which will be discussed later.

We first train the hybrid QCPN to predict the zeroth and first orders of the Bessel function, $J_0(x)$ and $J_1(x)$, as well as the fifth and sixth orders of the Legendre function, $P_5(x)$ and $P_6(x)$. The first hidden layer of the hybrid QCPN has two neurons generated from two five-qubit quantum circuits. For comparison, we also train a three-layer neural network (NN) to predict these functions. The architecture of the NN is shown in Fig.~\ref{Fig:fig5}(c), where the first layer employs the Tanh activation function, and the second layer uses a linear function. Both networks are trained using 50,000 samples and tested with 1,000 samples. Figures~\ref{Fig:fig5}(d)-~\ref{Fig:fig5}(g) show the predictions from both models. The red symbols represent the prediction of the hybrid QCPN, and the blue symbols indicate the prediction of the NN, with the dashed curve representing the exact values of these special functions. The hybrid QCPN model achieves high prediction accuracy for all four functions. In contrast, the NN model fails to accurately predict $J_0$ and $J_1$ as $x>5$ and $P_5(x)$ and $P_6(x)$. Increasing the number of neurons in the second layer of the NN does not improve accuracy, whereas adding one or two more layers is necessary for the NN to succeed. These results demonstrate that the hybrid QCPN, with its higher-order nonlinear activation functions, can reduce the depth of a neural network, thereby decreasing the overall model complexity.

Furthermore, we examine Eq.~(\ref{eq: multchev}) by training the hybrid QCPN on four additional special functions, $z=J_0(x+y)$, $z=J_1(x+y)$, $z=P_5(xy)$, and $z=P_6(xy)$. The curves of these special functions are shown in Figs.~\ref{Fig:fig5}(h)-~\ref{Fig:fig5}(k). For each QCPN, we use a five-qubit configuration, while the classical layer consists of ten input nodes. The red, green, and blue circles in Figs.~\ref{Fig:fig5}(h)-~\ref{Fig:fig5}(k) represent the predicted values of the trained hybrid QCPN for the testing dataset. The model achieves high accuracy in predicting these special functions, indicating that the hybrid QCPN can effectively tackle complex machine-learning tasks involving input with multiple elements.

Similarly, this method can be adapted to classical neural networks to enhance their performance. A related approach was recently proposed in the context of Chebyshev polynomial-based Kolmogorov-Arnold Networks~\cite{ss2024chebyshev}. Unlike the method in Ref.~\cite{ss2024chebyshev}, the QCPN can automatically express a continuous function using a polynomial with multiple terms, eliminating the need to construct Chebyshev bases manually. In addition, the QCPN can build a Chebyshev function up to the $(n-1)(n-2)$-th order using only $n$ qubits, which is more efficient than the classical method. Furthermore, we utilize a different polynomial function to describe continuous functions with high-dimensional inputs.

\section{Perspective}\label{sec:IV}
Our work investigates the functional expressibility of quantum circuits for machine learning tasks, providing insights distinct from those in Ref.~\cite{cerezo2024does}, which primarily explores the capacity of simulating the quantum state of a quantum circuit. Unlike classical neural networks, quantum neural networks leverage specialized activation functions that can potentially reduce both the number of training steps and the model size. Interestingly, similar benefits can be achieved on classical computers by employing quantum-inspired activation functions. This finding raises an important question: Do we really need to use quantum computers to achieve these benefits?

The answer is yes! Here, we aim to discuss the benefits of leveraging quantum circuits and quantum computers for machine-learning tasks. In our work, we manually design a quantum-inspired activation function to extract features from input images. This construction is relatively straightforward when the quantum circuit is small. However, as the size of the quantum model scales to hundreds or thousands of qubits, constructing an efficient polynomial function for machine learning tasks becomes increasingly challenging. Adopting a quantum circuit can significantly simplify the process of building such polynomial functions. For instance, one can repeatedly apply specific quantum layers within a quantum circuit to create complex quantum activation functions. However, developing such large-scale quantum models depends on advances in quantum computing hardware, which would require highly reliable quantum computers with at least 1,000 logical qubits. Recent developments, such as IBM’s error-correction techniques utilizing fewer physical qubits, have shown promise in realizing quantum computers with a large number of logical qubits~\cite{High2024Bravyi}.

In addition, developing quantum machine learning techniques is crucial for addressing the challenges inherent in quantum systems. For example, a quantum machine learning model can be directly employed to mitigate quantum errors in quantum computers~\cite{Ohno2023direct}. Moreover, quantum machine learning models can be used to classify the quantum states of a system simulated on quantum computers~\cite{Uvarov2020Machine}. These applications highlight the importance of quantum machine learning techniques in solving quantum many-body physics problems. 

Understanding the mathematical expression of quantum circuits also motivates us to explore quantum-inspired activation functions. As we discussed, constructing an efficient polynomial function with hundreds or thousands of terms is challenging. It would be interesting to explore some rules to guide us to construct polynomial functions with high learning capability. If such a rule can be derived, developing a quantum model for classical machine learning tasks would become less critical.

\section{Conclusion}\label{sec:V}
Our work explored the functional expressibility of a generic quantum circuit. By examining two shallow quantum circuits integrated into a convolutional neural network, we demonstrated that incorporating a quantum activation function can enhance model performance by reducing the required number of training steps. Additionally, we investigated three quantum-inspired activation functions and found that they can achieve prediction accuracies comparable to those of standard classical CNNs. Inspired by the properties of Chebyshev polynomial functions, we developed a hybrid quantum Chebyshev-polynomial network capable of approximating any continuous function using just three layers. In contrast, a standard three-layer neural network fails to predict continuous functions with high-order nonlinearities.

We provided an in-depth discussion on future research directions in quantum machine learning. We advocate for efforts to optimize large-scale quantum circuits and to explore systematic methods for constructing efficient polynomial functions with thousands of terms. Advancements in either of these areas could significantly enhance machine learning model performance by reducing the number of training steps and model complexity. This is particularly crucial for large-scale natural language models like ChatGPT and Gemini.

\section*{Acknowledgement}
This work is supported by the National Center for Transportation Cybersecurity and Resiliency (TraCR) (a U.S. Department of Transportation National University Transportation Center) headquartered at Clemson University, SC, USA. Any opinions, findings, conclusions, and recommendations expressed in this material are those of the author(s) and do not necessarily reflect the views of TraCR. The U.S. Government assumes no liability for the contents or use thereof. We acknowledge the computational support from Palmetto, a supercomputing cluster at Clemson University.

%This work is supported by the National Center for Transportation Cybersecurity and Resiliency (TraCR) (a U.S. Department of Transportation National University Transportation Center) headquartered at Clemson University, Clemson, South Carolina, USA. Any opinions, findings, conclusions, and recommendations expressed in this material are those of the author(s) and do not necessarily reflect the views of TraCR. The U.S. Government assumes no liability for the contents or use thereof. We acknowledge the computational support from the Palmetto, a high-performance computing cluster at Clemson University.

%\section*{Conflict of Interest}
%None of the authors has a conflict of interest of disclose.

\begin{figure*}[ht]
\centering
\includegraphics[width=0.7\textwidth]{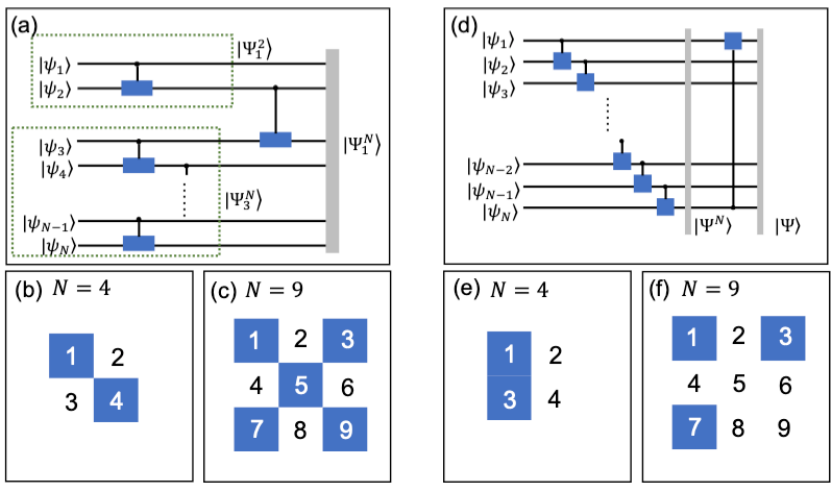}
\centering\caption{(a) The quantum state $|\Psi_1^N\rangle$ of quantum circuit 1 is divided into two parts, $|\Psi_1^2\rangle$ and $|\Psi_3^{N}\rangle$, where $N$ is the total number of qubits. (b) The contribution of a $2\times 2$  filter encoded into quantum circuit 1 comes from the first and fourth qubits. (c) The contribution of a $3\times 3$ filter encoded into quantum circuit 1 comes from qubits with odd indexes. (d) The analysis of quantum circuit 3. (e) The contribution of a $2\times 2$  filter encoded into quantum circuit 3 comes from the first and third qubits. (f) The contribution of a $3\times 3$ filter encoded into quantum circuit 3 comes from the first, third, and seventh qubits. 
}
\label{Fig:fig6}
\end{figure*}

%\appendices
\section*{Appendix}
\subsection*{The function expression of quantum circuit 1}

To obtain the output $\langle \Psi | Z^{\otimes N}|\Psi \rangle$ of QC 1, we need to use the following proposition. Here, $|\Psi\rangle$ is the final quantum state of this quantum circuit. 
\begin{Proposition}
We consider a quantum circuit composed of $N$ qubits, where the initial state $|\psi_i\rangle$ of each qubit is described by $|\psi_i\rangle=q_i|0\rangle + p_i | 1\rangle$. Upon the application of a CNOT gate between two adjacent qubits first across even-indexed qubits and then odd-indexed qubits, as depicted in Fig.~\ref{Fig:fig6}(a), 
the expectation value of the operator $Z^{\otimes N}$ is determined by the following recursive relationship:
\begin{equation}
\langle \Psi_1^N | Z^{\otimes N} | \Psi_1^N \rangle = \left[ |p_1|^2-|q_1|^2  \right] \times \langle \Psi_3^{N}| Z^{\otimes N-2} | \Psi_3^{N}\rangle,\nonumber\\
\end{equation}
where $|\psi_n^m\rangle =\otimes_{i=n}^{i=i+2,i<m}\text{CX}(i,i+1)\otimes_{j=n+1}^{j=j+2,j<m}\text{CX}(j,j+1)\otimes_{i=n}^m |\psi_i\rangle$.
The notation $\text{CX}(i,j)$ denotes a CNOT gate with qubit $i$ as the controller and qubit $j$ as the target.
\end{Proposition}

To demonstrate this proposition, we treat the first two qubits and the remaining qubits, shown in Fig.~\ref{Fig:fig6}(a), separately. The quantum state on the first two qubits after applying a CNOT gate is
\begin{equation}
|\Psi_1^2\rangle\!=\!q_1q_2|00\rangle \!+\! q_1p_2|01\rangle + p_1q_2|11\rangle \!+\! p_1p_2|10\rangle.
\end{equation}
We label the quantum state on the remaining $N\!-\!2$ qubits as $|\Psi_3^{N}\rangle$ and entangle this state with $|\Psi_1^2\rangle$ via a CNOT gate $\text{CX}(2,3)$. The entangled quantum state is then given by
\begin{align}
|\Psi_1^N\rangle \!=\!& \big[ q_1 q_2 |00\rangle + q_1 p_2 |01\rangle X_3 + p_1q_2|11\rangle X_3 + p_1p_2|10\rangle   \big]  \nonumber\\
& \otimes |\Psi_3^N\rangle. 
\end{align}
Here, $X_3$ is the $x$-component Pauli matrix that acts on the third qubit. Utilizing the relationship that $\langle \Psi_3^N|X_3 Z^{\otimes N-2}X_3 | \Psi_3^N\rangle\!=\!-\langle \Psi_3^N|Z^{\otimes N-2}| \Psi_3^N\rangle$, we find
\begin{equation}\label{eq:psiNZpsiN}
\langle \Psi_1^N|Z^{\otimes N} |\Psi_1^N \rangle \!=\! \left[ |p_1|^2 \!-\! |q_1|^2 \right] \langle \Psi_3^N|Z^{\otimes N-2} |\Psi_3^N \rangle.
\end{equation}

Proposition 1 is demonstrated from the above derivations. With proposition 1, we can easily obtain the output of QC 1. If the number of qubits $N$ is odd, the output is $\langle \Psi_1^N | Z^{\otimes N} | \Psi_1^N \rangle\!=\!-\prod^N_{i\in \text{odd}}\left[ |p_i|^2 -|q_i|^2 \right]$; if $N$ is even, then the output becomes $\langle \Psi_1^N | Z^{\otimes N} | \Psi_1^N \rangle\!=\!-\left[ |p_N|^2 -|q_N|^2 \right]\times \prod_{i\in \text{odd}}^{N-3}\left[ |p_i|^2 -|q_i|^2 \right]$. In QC 1, $p_i$ and $q_i$ depend on the values of $\theta_i$ and $\phi_i$. With some effort, we find that the output of QC 1 can be written as
\begin{equation}\label{eq:QC1}
O_{QC1}(\theta,\phi)=\left\{ \begin{matrix}
-\prod^N_{i\in\text{odd}} \text{cos}\alpha_i & N\in\text{odd} \\
-\text{cos}\alpha_N\prod_{i\in\text{odd}}^{N-3} \text{cos}\alpha_i & N\in\text{even}
\end{matrix},
\right.
\end{equation}
where $\alpha_i\!=\!\theta_i+\phi_i$. Equation~(\ref{eq:QC1}) shows that the output of QC 1 depends on rotation angles on specific qubits. For example, the output of a $2\times 2$ quantum filter depends on the angle on the first and fourth qubits, which are highlighted by the blue color in Fig.~\ref{Fig:fig6}(b). For a $3\times 3$ filter, the output is determined by the angle on the qubit with odd indexes (see Fig.~\ref{Fig:fig6}(c)).

\subsection*{The function expression of quantum circuit 2}
Our investigation into the activation function of QC 2 commences by examining the quantum state prior to the final CNOT gate application, denoted as $|\Psi^N\rangle$, and illustrated in Fig.~\ref{Fig:fig6}(d). Proposition 2 provides the necessary framework for this analysis, establishing that $|\Psi^N\rangle$ can be succinctly described through a recursive formulation, detailed as follows.
\begin{Proposition}
Upon the application of a CNOT gate between adjacent qubits within a quantum circuit starting from the initial state $|\psi_i\rangle$ of each qubit, the resultant quantum state evolves into
\begin{align}\label{eq:psiN}
&|\Psi^N\rangle = \left[ p_N|g^{N-2}\rangle\otimes|0\rangle_{N-1} + q_N|t^{N-2}\rangle\otimes|1\rangle_{N-1}  \right]\otimes \nonumber\\
&|0\rangle_N + \left[ q_N|g^{N-2}\rangle\otimes|0\rangle_{N-1} +  p_N|t^{N-2}\rangle\otimes|1\rangle_{N-1}    \right]\otimes|1\rangle_N,
\end{align}
where $|g^{N-2}\rangle$ and $|t^{N-2}\rangle$ represent states across the first $N-2$ qubits. These two quantum states follow recursion relationships, which are expressed as
\begin{align}\label{eq:propos2}
|g^{N-2}\rangle =p_{N-2}|g^{N-4}\rangle\otimes|0\rangle_{N-3} + q_{N-2}|t^{N-4}\rangle\otimes|1\rangle_{N-3},&\nonumber\\
&\\
|t^{N-2}\rangle = q_{N-2}|g^{N-4}\rangle\otimes|0\rangle_{N-3}+p_{N-2}|t^{N-4}\rangle\otimes|1\rangle_{N-3}.&\nonumber\\ 
\end{align}
\end{Proposition}

Utilizing the relationship between the output wavefunction $|\Psi\rangle$ of QC 2 and $|\Psi^N\rangle$ and Eq.~(\ref{eq:psiN}), the output $\langle \Psi | Z^{\otimes N}|\Psi \rangle$ of QC 2 can be represented by $|g^{N-2}\rangle$ and $|t^{N-2}\rangle$ via 
\begin{equation}
\langle \Psi| Z^{\otimes N} | \Psi \rangle = \langle g^{N-2} | Z^{\otimes N-2} | g^{N-2} \rangle - \langle t^{N-2} | Z^{\otimes N-2} | t^{N-2} \rangle.
\end{equation}
Therefore, to obtain the result of $\langle \Psi| Z^{\otimes N} | \Psi \rangle$, we first need to evaluate $\langle g^{N-2} | Z^{\otimes N-2} | g^{N-2} \rangle$ and $\langle t^{N-2} | Z^{\otimes N-2} | t^{N-2} \rangle$. It is found that $\langle g^{N-2} | Z^{\otimes N-2} | g^{N-2} \rangle$ and $\langle t^{N-2} | Z^{\otimes N-2} | t^{N-2} \rangle$ follow the following two recursion equations,
\begin{align}
&\langle g^{N-2} | Z^{\otimes N-2} | g^{N-2} \rangle + \langle t^{N-2} | Z^{\otimes N-2} | t^{N-2} \rangle = \nonumber\\
&\langle g^{N-4} | Z^{\otimes N-4} | g^{N-4} \rangle - \langle t^{N-4} | Z^{\otimes N-4} | t^{N-4} \rangle,
\end{align}
and 
\begin{align}
&\langle g^{N-2} | Z^{\otimes N-2} | g^{N-2} \rangle - \langle t^{N-2} | Z^{\otimes N-2} | t^{N-2} \rangle = \nonumber\\
&\mathrm{cos}\alpha_{N-2}  \left[ \langle g^{N-4} | Z^{\otimes N-4} | g^{N-4} \rangle + \langle t^{N-4} | Z^{\otimes N-4} | t^{N-4} \rangle   \right]. \nonumber\\
\end{align}
Using these recursion equations, the output of QC 2 can be formalized as
\begin{widetext}
\begin{equation}
O_{QC2}(\theta,\phi) = \left\{ \begin{matrix}
\left(\prod_{i=N-2}^{i=i-4,i>4}\text{cos}\alpha_{i}\right)\text{cos}\alpha_3\text{cos}\alpha_1 &  N=4+4m\nonumber\\
\left(\prod_{i=N-2}^{i=i-4,i>6}\text{cos}\alpha_{i}\right)\text{cos}\alpha_4\text{cos}\alpha_2 & N=6+4m\nonumber\\
\left(\prod_{i=N-2}^{i=i-4,i>5}\text{cos}\alpha_{i}\right)\text{cos}\alpha_5\text{cos}\alpha_2 &  N=3+4m\nonumber\\
\left(\prod_{i=N-2}^{i=i-4,i>5}\text{cos}\alpha_{i}\right)\text{cos}\alpha_3\text{cos}\alpha_1 &  N=5+4m\nonumber,
\end{matrix} \right.
\end{equation}
\end{widetext}
where $n$ is a non-negative integer. Notably, $O_{QC2}$ also has a qubit index dependence. If we use a $2\times 2$ quantum filter, only rotation angles on the first and third qubits contribute to the output, highlighted as blue color in Fig.~\ref{Fig:fig6}(e). For a $3\times 3$ quantum filter, the output is determined by the rotation angle on the first, third, and seventh qubits (see Fig.~\ref{Fig:fig6}(f)).

%\bibliographystyle{IEEEtran} 
%\bibliography{reference}

\end{document}